\documentclass[aps,showpacs,prd,twocolumn]{revtex4-1}
 \usepackage{amsmath}
\usepackage{amssymb}
\usepackage{xcolor}
\usepackage{epsfig}
\usepackage{natbib}
\usepackage{epstopdf}
\usepackage{graphicx}
\usepackage{bm}

\usepackage[caption=false,font=Large]{subfig}

\newcommand*{\be}{\begin{equation}}
\newcommand*{\ee}{\end{equation}}
\newcommand*{\bea}{\begin{eqnarray}}
\newcommand*{\eea}{\end{eqnarray}}

\usepackage[svgnames]{xcolor} 
\usepackage{xcolor}

\definecolor{darkblue}{rgb}{0.0, 0.0, 0.55}

\definecolor{myblue}{rgb}{.2, .36, .77}

 \DeclareFontFamily{OT1}{pzc}{}
 \DeclareFontShape{OT1}{pzc}{m}{it}%
 {<->  s  *  [1.400]  pzcmi7t}{}
\DeclareMathAlphabet{\mathscr}{OT1}{pzc}%
{m}{it}

\allowdisplaybreaks

\begin{document}

\title{Dispersion  of  three-dimensional  oscillons:    a standing-wave perspective}

 \author{  I.\,V.\ Barashenkov$^*$,     R. De Sousa$^\dagger$,  N. V. Alexeeva$^*$,     A.\ Dika$^{\ddag}$,     E. V. Zemlyanaya$^{\P}$                 }
   
\affiliation{
$^*$ Centre for Theoretical and Mathematical Physics, University of Cape Town, 
South Africa    \\
$^\dagger$  Rudolf Peierls Centre for Theoretical Physics, University of Oxford,  
 UK   \\
$^\ddag$ School of Software Design and Data Science, Seneca Polytechnic,  Toronto,  Canada \\
$^{\P}$      M G  Mescheryakov 
Laboratory for Information Technologies, 
Joint Institute for Nuclear Research,  Dubna,  Russia
}

\begin{abstract}

A spherically symmetric standing wave confined within a finite-radius ball, with weak radiative exchange between its core and boundary, 
provides a model for an intermediate stage in the adiabatic evolution of the $\phi^4$ oscillon. 
We employ a multiscale variational framework to derive an approximate description of these standing waves and their energy–frequency relation. 
The resulting dispersion curve is compared with that obtained from numerical solutions of the boundary value problem for the $\phi^4$ equation on a two-dimensional (time and radius) domain. The agreement between the numerical and variational results is excellent near the  endpoint of the continuous spectrum but deteriorates 
as  $\omega$ is reduced below $0.85 \, \omega_0$. We attribute this discrepancy to the growth of the radiation wing, which is neglected in the variational ansatz.
\end{abstract}

\maketitle

\section{Introduction}

Oscillons are 
   spherically-symmetric solutions of the $\phi^4$ equation,
   \be
\Phi_{tt} -\Delta \Phi - \Phi + \Phi^3=0,
\label{A1}
\ee   
describing pulsating structures with 
 an exponentially localized core. Outside the core, the field remains close to one of its two vacuum states.

The field oscillations in the core generate outgoing radiation at the second and higher harmonics. 
 The continual loss of energy precludes exact periodicity; as a result, the oscillons are transient states with a  finite lifetime. 
 Their period of existence, however,  includes an extended epoch during which the
       energy loss is slow, the oscillations remain very close to periodic, while the structure  exhibits remarkable stability  to perturbations.
       Consequently,  these  states should play a role in all phenomena modelled within the $\phi^4$ and related theories. 
      
Oscillons (known as {\it pulsons\/}  at the time) were first identified in the numerical simulations of Ref.~\cite{BM2}, 
which were motivated by the study of coherent structures  in nonlinear wave systems.
The widespread interest in these objects was sparked by Ref.~\cite{G1}, which highlighted their relevance across a range of cosmological scenarios, including  symmetry-breaking phase transitions and false vacuum decay 
\cite{CGM, Riotto, GInt, 11cosmo, Dymnikova, Broadhead, bubbling, Amin1, Stamatopoulos, Zhou, Amin2, Adshead, GG}.
 Since their discovery, oscillons have  been shown to arise in a wide variety of field theories, including the planar Abelian Higgs model \cite{GT2,Achi}, 
 the Einstein–Klein–Gordon equations \cite{Maslov, Zhang2, Kou1}, 
 string-motivated scenarios \cite{string,Kasu,Sang},  axion frameworks \cite{Kolb,Vaquero,Kawa_axion,Olle,Miyazaki,Cyn}, 
and the bosonic sector of the Standard Model \cite{Farhi2,Graham,Gleiser4,Sfakianakis,Correa}. 
The quantum dynamics of oscillons  was studied  in \cite{Hertz,Saffin}, and  fermionic degrees of freedom were included  in \cite{Borsanyi}.

 Recent work has focused on the emergence of oscillons in the early Universe, where they are expected to form through inflaton self-resonance and fragmentation during the preheating stage following inflation. Refs.~\cite{Hong,LozAm,Mahbub,Kim,Kou2,Shafi} investigate the conditions for fragmentation, the dependence of oscillon formation on the inflaton potential, and the subsequent nonlinear evolution of the scalar field. An oscillon-dominated   period can have significant cosmological consequences, including gravitational-wave production, transitions between matter- and radiation-dominated expansion, and possible connections with primordial black-hole formation \cite{Antusch,Drees,Hira,Piani_2025,Sui,Lozanov_2026,Ballesteros}.

 More broadly, oscillons have been studied as  long-lived, localized nonlinear structures, with their properties determined by the underlying field theory and by the interplay between scalar self-interactions, radiation, gravity, and couplings to additional degrees of freedom. This perspective has been explored in a variety of field-theoretic settings, including models with gravity and multiple interacting fields. In particular, numerical-relativity studies have investigated the conditions under which dense oscillons remain quasistable, disperse, or undergo gravitational collapse to form black holes \cite{Nazari,Kou2}. Ref.~\cite{Shafi} shows that additional scalar couplings can modify oscillon formation and introduce new decay channels. Higgs-like and scalar--tensor models provide further examples in which quartic interactions and additional field couplings strongly influence oscillon formation and lifetime \cite{Piani,Piani_2025}.

{
Computer simulations \cite{G1,CGM,Honda,Gleiser10} and Fourier-based numerical analyses \cite{Honda,Fodor1,Cyn} have remained the primary tools for investigating oscillon formation and properties.
 Alongside the numerical work, analytical approaches have been developed to estimate key oscillon characteristics, 
 including energy, radius, frequency, core amplitude, and lifetime \cite{GS1,GS2}. Those estimates rely on a heuristic combination of linear radiation theory and a single-mode variational approximation \cite{CGM,G2,GS1,GS2,Pujo2}.   
 A distinct analytical approach was pursued in Ref.~\cite{Fodor_Mezei}, where the radiation loss of small-amplitude oscillons was evaluated using asymptotics beyond all orders and Borel summation.  A truncated sequence of coupled Klein-Gordon equations for the Fourier modes was examined in \cite{Atanas}.
  
  Here, we study solutions of equation \eqref{A1} that are closely related to oscillons but possess more tractable properties \cite{ABBZ}. 
    These solutions represent  standing waves  that form  due to the interference of 
    the outgoing radiation  and radiation  reflected from the walls of a finite-size cavity.
    Mathematically, they are strictly time-periodic   configurations  satisfying $\Phi(R,t)=\Phi_0$, where $\Phi_0$
    is one of the two vacuum states and $R$ denotes the radius of the confining ball. 
    Standing waves belong to the same periodic family as nanopterons \cite{Boyd} or quasibreathers \cite{Fodor1,Cyn,Fodor2,Nagy}, but are supported by a bounded spherical domain.   
       Considering periodic structures enclosed within a finite-radius cavity endows these configurations with a clear physical interpretation and,
    at the same time, 
     introduces a 
    quantitative parameter distinguishing between solutions sharing the same period.

Our interest in the standing waves stems from their utility as approximations to the oscillons. 
At the stage of weak energy loss, the oscillon performs an adiabatic drift through phase space, with its parameters remaining constant over long time intervals. Throughout this adiabatic drift, the oscillon stays close to a standing wave with a thin wing of outgoing and reflected radiation \cite{Fodor1,ABDD}.

 In this work, 
we investigate standing waves from two complementary perspectives, numerical and variational. Numerically, periodic solutions of equation \eqref{A1} are obtained by solving a boundary-value problem on the two-dimensional domain $(0,T)\times(0,R)$;  this approach allows us to construct the dispersion curve over a wide range of frequencies. Complementing this, we formulate a multiscale variational framework, which offers analytical insight into selected features of the dispersion curve.
 
 The numerical  energy-frequency diagram of the standing wave is reported in section \ref{overview}
 while our variational formalism is laid out in 
 section \ref{variational}.
 Section 
 \ref{fixed} summarises the analysis of the  variational fixed points, with the details of the calculation
 relegated to the Appendices. 
 Finally, section \ref{conclusions} draws conclusions from the numerical-variational comparison.

  \section{Boundary-value problem on a cylindrical domain}
\label{overview}

      In what follows, we fix the vacuum state to be $\Phi_0=-1$ for definiteness.   The difference
\[
\phi=  \Phi-\Phi_0
\]
obeys
\begin{subequations}  
\label{2D}
\be
\phi_{tt}- \phi_{rr} - \frac{2}{r} \phi_r + 2 \phi- 3 \phi^2 + \phi^3=0.
\label{A200}
\ee
As in \cite{ABBZ} we determine standing waves as solutions of a 
 boundary-value problem on a two-dimensional domain.
The boundary conditions are
\be
\label{BC}
\phi_r(0,t)= \phi(R,t)=0
\ee
and
\be
\phi(r,    T)=\phi(r,0).
 \label{BCT}
\ee
\end{subequations}    

The confining ball radius, $R$, and the period $T$ are two parameters that determine the solution of the boundary-value problem \eqref{2D}. 
The frequency of the standing wave is defined by $\omega= 2 \pi/T$.
In what follows, we measure frequency in units of $\omega_0$,    the endpoint of the continuous spectrum of vacuum excitations
(the mass of elementary constituents of the model). With the normalisation adopted in equation \eqref{A1}, the endpoint is given by  
\be
\omega_0 = \sqrt 2.
\label{ome0}
\ee
Measuring the  frequency in units of $\omega_0$  facilitates comparison with results 
obtained by authors who use different   coefficients in their versions of Eq. \eqref{A1}.

In Ref.~\cite{ABBZ}, the boundary-value problem \eqref{2D} was studied for $\omega/\omega_0$ in the interval $(0.85,1]$. 
Here, we extend the standing-wave solution to a wider frequency interval $(0.48,1]$, while keeping the ball radius fixed at $R=100$.


  \begin{figure}
      \includegraphics[width= 0.9 \linewidth]{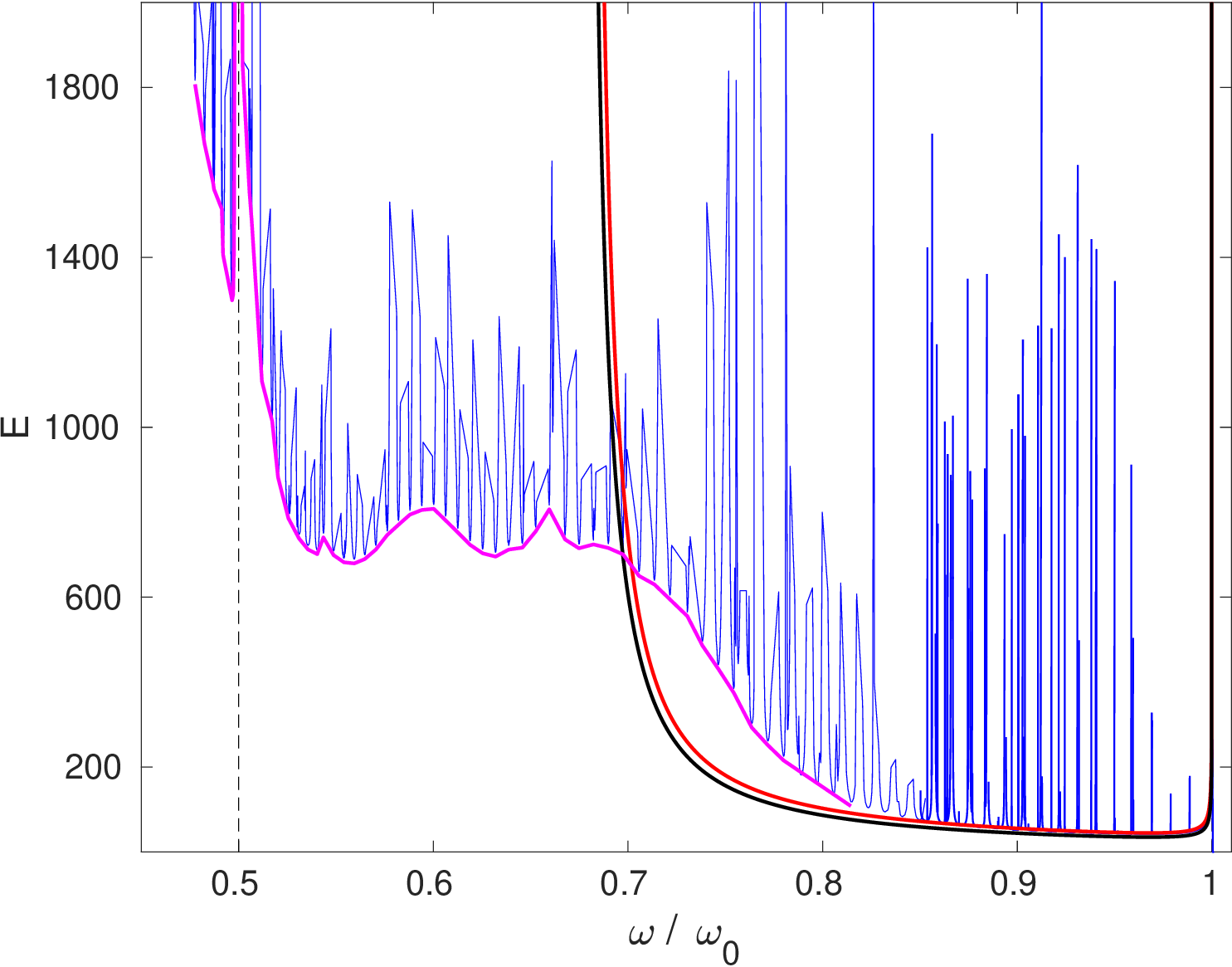}   
      
       \vspace*{2mm}
         \includegraphics[width=0.9 \linewidth]{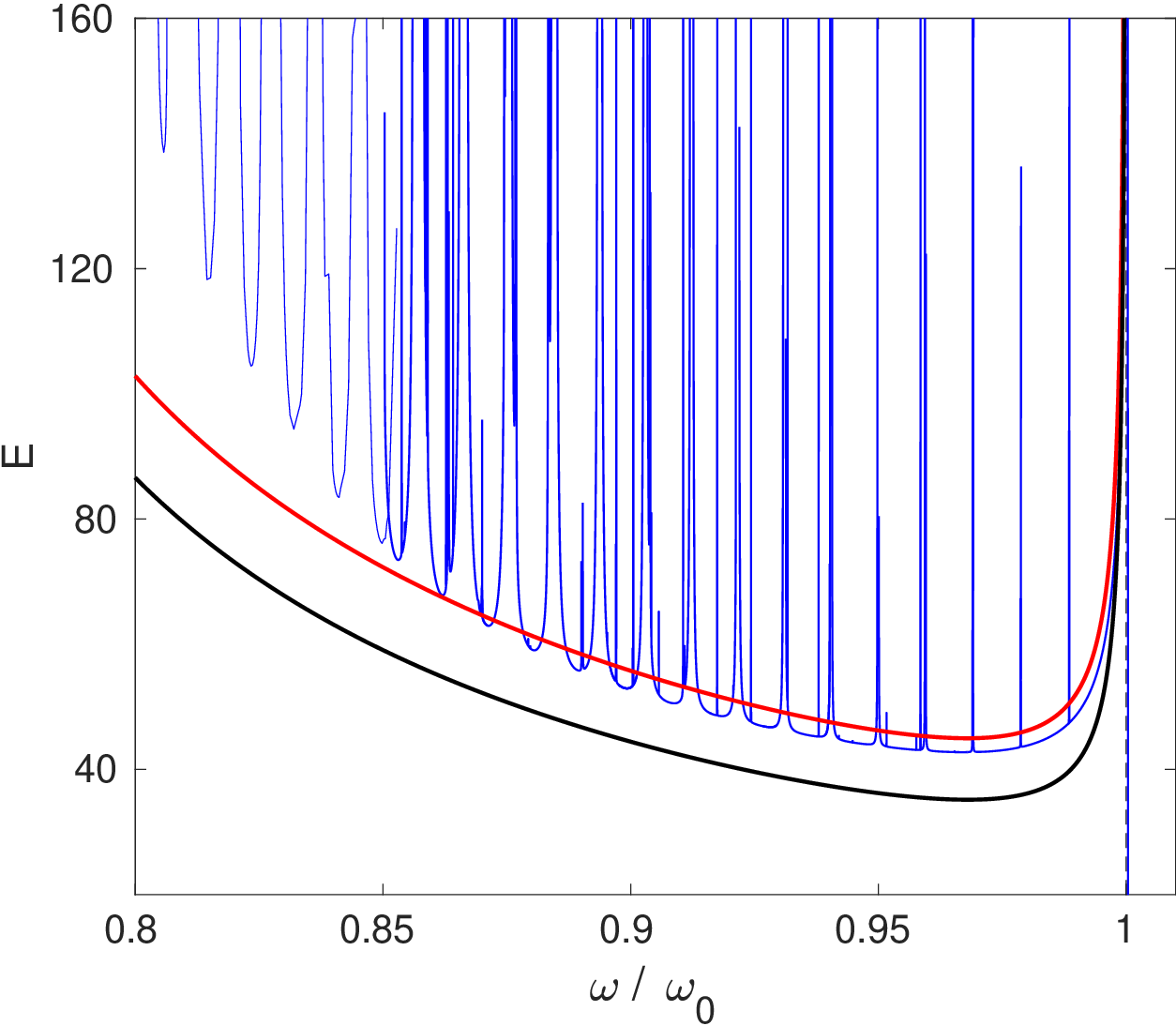}    
            \caption{ (a): Dispersion of three-dimensional standing waves. Blue:  energy of the numerical  solution of the boundary-value problem \eqref{2D}. 
   Pink: lower envelope of the energy minima, which is expected to lie close to the envelope of the family of dispersion curves obtained for varying $R$. 
   Red: energy of the fixed point of the variational equations, given by Eq.~\eqref{EU}. The minimum of the variational energy is at $\omega_{\mathrm{min}}=0.96795 \, \omega_0$.
   Black: energy of the variational fixed point with $F=0$. (b): Blow-up of the segment containing the minimum of the envelope.} 
 \label{En1} 
\end{figure}

A physically meaningful characteristic of 
standing waves  is their energy,
  \be
E[\phi]=  4 \pi\int_0^R\left( \frac{\phi_t^2}{2} +\frac{\phi_r^2}{2} + \phi^2-\phi^3+\frac{\phi^4}{4} \right) r^2 dr. 
\label{E}
\ee
Energy is the only conserved quantity 
of equations \eqref{2D}.
Of special importance is  the {\it dispersion\/} of the localised solution, that is, its energy-frequency relationship. 
Whether an analytical model captures the essentials  of the standing wave’s structure and dynamics 
can be assessed from the agreement between the analytical and numerical dispersion curves.

The localised solutions of the boundary-value problem \eqref{BC} are {\it nanopterons}. 
As \(r\to R\), the field \(\phi\) does not decay monotonically to zero. 
Instead, an exponentially localised core is followed by an oscillatory tail, or wing,  that extends to the boundary of the ball, with its amplitude decaying as \(1/r  \).

Fig \ref{En1} displays the numerically obtained dispersion curve for the standing waves in a ball of radius $R=100$. The curve was computed using Newton iteration with second-order finite-difference approximations for both the radial and temporal derivatives. 
The radial and temporal discretisation stepsizes were $\Delta r=100/512$ and $\Delta t=T/1000$, respectively.

   As $\omega$ is varied, the energy $E[\phi_{\omega, R}]$ 
  develops a sequence of resonant peaks.  The peaks result from the thickening  of the radiation  wings
  (that is, from the   growth of the  amplitude of the slowly decaying part 
  of solution external to its core). The positions of the peaks are determined by the radius of the ball \cite{ABBZ,ABDD}.
 Varying $R$ over a narrow interval $(100-\lambda/2, 100+\lambda/2)$, where 
 $\lambda=2\pi/k$  and $k=\sqrt{(2\omega)^2-\omega_0^2}$ is the radiation wavenumber, produces a dense family of jagged dispersion curves bounded from below by a continuous envelope.
The envelope is traced by a pink line in Fig \ref{En1}. 
 
Unlike its one-dimensional counterpart \cite{ABDD}, the envelope of the 3D energy curve in Fig.~\ref{En1} does not decrease monotonically with increasing frequency. In fact, $E(\omega)$ blows up as $\omega\to\omega_0$; this divergence can be accounted for by expanding the wave core in powers of $\sqrt{\omega_0^2-\omega^2}$ \cite{Voronov2,Fodor2} or, alternatively, in powers of $R^{-1}$ \cite{ABBZ}. 

The energy envelope features a minimum at $\omega_{\mathrm{min}}=0.96732   \,   \omega_0$, which will reappear in our variational analysis below. The pronounced undulations of $E(\omega)$ for $\omega\lesssim   \, 0.7 \,  \omega_0$ point to increasingly complex spatio-temporal structure of the solutions in that frequency range.

We note the discontinuity of the envelope of the $E(\omega)$ curve that occurs at $\omega = \omega_0/2$.
When $\omega$ exceeds $\omega_0/2$, its overtone $2\omega$ falls within the continuous spectrum, opening up the second-harmonic radiation channel in addition to the third- and higher-harmonic bands that are already available to cores oscillating at frequencies $\omega < \omega_0/2$.  The second-harmonic radiation waves have larger amplitudes than their third-harmonic counterparts; this explains the energy hike observed as $\omega/\omega_0$ is raised through $1/2$ \cite{ABDD}.

The dispersion discontinuity has a common origin with the radiation emitted in staccato flashes by a slowly decaying oscillon in certain one-dimensional Klein–Gordon equations \cite{Bogolubsky_PL,Shnir,Nagy}.

The classification of stability of the standing-wave solutions represented by the diagram in Fig. \ref{En1} is beyond the scope of the present study. At this stage, the only available information on stability comes from earlier work \cite{ABBZ}, which found a pair of real Floquet multipliers outside the unit circle throughout the entire range $\omega > \omega_{\mathrm{min}}$. By contrast, the Floquet spectrum for $\omega < \omega_{\mathrm{min}}$ remains largely unexplored. Here, we merely note that the unstable spectrum in that region should include both modes localised in the wave core  (like those found for $\omega_{\mathrm{min}} < \omega < \omega_0$)
 and modes associated with the radiation wings.

  \section{Variational formalism}
  \label{variational}

\subsection{Two time scales}

The variational approximation aims to capture the mechanism underlying the formation and persistence of the localised standing wave through a small set of 
collective coordinates. Its fidelity can then be assessed by comparing the resulting dispersion curve with the numerical one.

   As in \cite{BA,ABDD}, we introduce the two time scales ${\mathcal T}_0=t$ and ${\mathcal T}_1=\epsilon t$ and regard $\Phi$ as a function of both.
   The dependence on the fast time ${\mathcal T}_0$ is taken to be periodic with period $T$,
\[
\Phi(\mathcal T_0+T;     \, \mathcal T_1)=\Phi(\mathcal T_0;     \, \mathcal T_1),
\]
with the variation with respect to either time variable assumed to be of order one:  $\partial \Phi/ \partial \mathcal T_0, \, \partial \Phi/ \partial  \mathcal T_1 \sim 1$.
   Treating $\mathcal T_0$ and $\mathcal T_1$ as independent variables, the Lagrangian becomes
   \begin{align} 
L= 4\pi
 \int \left[   \frac12\left( \frac{\partial  \phi}{\partial {\mathcal T_0} } + \epsilon \frac{\partial  \phi}{\partial {\mathcal T_1} }   \right)^2
-\frac12\phi_r^2   
- \phi^2 \right.     \nonumber   \\ \left.   +\phi^3 -\frac14\phi^4 
 \right] r^2\ dr,
 \label{A2}
\end{align} 
where we let $\Phi=-1+\phi$.
    The action $\int L dt$ is replaced with
\be
\label{A3}
S=
 \int d   {\mathcal T_1} \, 
  \int_0^T  d   {\mathcal T_0}  \,
  L   \left(    \phi,  \frac{\partial  \phi}{\partial {\mathcal T_0} }, 
   \frac{\partial  \phi}{\partial {\mathcal T_1}}     \right).
\ee

To formulate the variational ansatz, 
it is instructive to recall 
the small-amplitude
asymptotic expansion  for the standing wave \cite{Voronov2,Fodor2}: 
\be 
\phi= 2 \varepsilon \mathcal R  ( \varepsilon \bm{ \xi} )   \cos (\omega t)  + \varepsilon^2 \mathcal R^2(\varepsilon \bm {\xi})   \left[3- \cos(2 \omega t) \right] + O (\mathcal \varepsilon^3 ),
\label{AA29}
\ee
where $\omega^2= \omega_0^2-\varepsilon^2$, $\varepsilon \to 0$  and $\mathcal R$ satisfies
\be
\nonumber
 \frac{d^2}{d\bm{\xi}^2}  
 \, \mathcal R- \mathcal R   + 12 \mathcal R^3=0.
\ee
(We distinguish between $\varepsilon$ in   \eqref{AA29} 
 and $\epsilon$ in  \eqref{A2}
  which denote two different small parameters.)
Note that the zeroth and second harmonic contributions to $\phi$ are quadratic in $\varepsilon$. This contrasts with the model with the $\mathbb{Z}_2$-symmetric vacuum \cite{Kose,BA}, where the next-to-leading harmonics are of order $\varepsilon^3$. Thus, unlike in the $\mathbb{Z}_2$-symmetric model, where retaining only the leading harmonic provided an accurate variational approximation \cite{BA}, the $\Phi^4$ equation requires the next-to-leading harmonics to be retained as well.
(The individual significance of the zeroth and second harmonics will be re-examined  in Section \ref{redundancy}
 below.)

The asymptotic expansion \eqref{AA29}, together with the one-dimensional variational ansatz of \cite{ABDD}, suggests the following form for the trial function:
   \begin{align}
\phi= A \,  \cos (\omega {\mathcal T_0} +\theta ) \, e^{-\left(r/b\right)^2}  
+ \left[  B \right. 
 \nonumber  \\  \left.
+ F   \cos 2( \omega {\mathcal T_0} +\theta )   \right]   e^{-2\left(r/b\right)^2} ,  
 \label{A4}
\end{align}
where $A, B, F,  b$ and $\theta$ are functions of the  ``slow'' time variable $\mathcal T_1$,
while $\omega = 2\pi/T$ ($\omega>0$).

After integrating over the explicit dependence on $r$ and $\mathcal T_0$, equations \eqref{A2} and \eqref{A3} yield
\be
S=       \frac{\pi^{3/2}}      {2}         \,     T \int d  {\mathcal T_1}  \, \mathcal L_\omega,
 \label{A5}
\ee
where the effective Lagrangian $\mathcal L_\omega$ is given by 
 \be
 \mathcal L_\omega = K
 +   
  \frac{ I}{2}  \left(\dot{\theta }+\omega \right)^2
  -U,
 \label{Lw}
 \ee
with
\begin{align}
K= \frac{b^3}{4} \left(  \frac{{\dot A}^2}{\sqrt 2}  + \frac{{\dot B}^2}{2}  + \frac{{\dot F}^2}{4}  \right) 
+   \frac{15        b {\dot b}^2           }{16}  \left( \frac{A^2}{ \sqrt 2}  +  \frac{B^2}{2}        \right.     \nonumber \\     \left. +   \frac{F^2}{4} \right)   
+ \frac{3         \dot b b^2           }{4} \left(
\frac{   \dot A A           }{ \sqrt 2}  +  \frac{ \dot B B}{2}   + \frac{\dot F F}{4} \right),
\label{A100}
  \end{align} 
  \be
   I = \frac{ b^3}{2 \sqrt 2}\left(
  A^2 +         \sqrt 2      F^2         \right),    
  \label{phi}
  \ee  
  and 
    \begin{align}
    U= \frac{3b}{4}  \left( \frac{A^2}{\sqrt 2} + B^2 + \frac{F^2}{2} \right) + \frac{b^3}{2} \left( \frac{A^2}{\sqrt 2}  
      + \frac{B^2}{2}   \right.  \nonumber \\   \left.  + \frac{F^2}{4} \right)  
    - \frac{b^3 B}{3 \sqrt 6} \left( B^2+ \frac32 F^2  \right)  -  \frac{3b^3A^2 }{8} \left(  B +\frac{F}{2} \right) \nonumber 
    \\
    + \frac{b^3}{4} \left[  \frac{3A^4}{32}  +  \frac{A^2}{\sqrt 6} \left( B^2+   BF +  \frac{F^2}{2} \right)    \right.   \nonumber  \\
    \left.  +  \frac{1}{8 \sqrt 2} \left(B^4 + 3 B^2F^2  +\frac38 F^4 \right)  \right].
 \label{A6}
  \end{align}
  In \eqref{A100}, we introduced a short-hand notation
  $\dot f \equiv \epsilon \partial f/  \partial  \mathcal T_1$.

  \subsection{Four degrees of freedom}

 The coordinate $\theta$ being cyclic  gives rise to the first integral 
\be
 \label{A700}   
\ell= (\dot{\theta}+\omega )   \, I,
\ee
where $I=I(A,F,b)$ is as in \eqref{phi}. 
Eliminating $\dot \theta$ by means of \eqref{A700} from the remaining equations, we arrive at the four-dimensional 
($A,B, F$ and $b$) system with  the 
Lagrangian 
 \begin{subequations}
 \label{Lagell}
 \be
 \label{LL}
  \mathcal L_\ell   = K  - U_\ell,
 \ee 
where
 \be
  U_\ell=   U+ \frac{\ell^2}{ 2  I}.     
  \label{UL}
   \ee
   \end{subequations} 
   
Note that the frequency has relinquished its role as an external parameter in favour of
 the quantity $\ell$. 
The {\it stationary\/}  solutions of the system \eqref{Lagell}
can still be characterised by  $\omega$, 
which can be recovered from $\ell$ and the components of the fixed point via
\be 
    \omega=\frac{\ell}{I}.          
\label{C1}
\ee 
(In fact, $\omega$ is a more natural parameter than $\ell$ in determining the fixed points;
see the next section.)
However, when considering  the {\it dynamics\/}  of the collective coordinates $A,B,F$ and $b$
--- for instance,   when classifying the stability of the fixed points ---  it is the 
parameter $\ell$ that has to be held fixed.

The system \eqref{Lagell}  conserves energy:
\be
\label{H}
H=   K+U_\ell.
 \ee
 This scalar characteristic serves as a bridge between the finite-dimensional dynamics and the full
 partial differential equation \eqref{A1}.
 To make this connection explicit, we relate the conserved quantity \eqref{H} to the field energy carried by the  trial function \eqref{A4}.

If a trial function $\phi (r,t)$ 
is not 
a solution of the equation \eqref{A200}, the associated 
field energy \eqref{E}
is not conserved.
However
 we may consider the 
{\it average\/} energy associated with this trial function:
 \be 
\label{E2}
{\overline E} =\frac{1}{T} \int_t^{t+T}  E[\phi(r, t^\prime)]  \, d t^\prime.
\ee
If  $\phi$ is 
$T$-periodic, $E[\phi]$  is also $T$-periodic and so 
 $\overline E$ is independent of time.

Substituting the ansatz \eqref{A4} --- with frequency $\omega$ and constant parameters $A,B,F,b$ --- into equations \eqref{E} and \eqref{E2} yields
\be
\overline E = \frac{\pi^{3/2}}{2} U_\ell(A,B,F,b)
= \frac{\pi^{3/2}}{2} H,
\label{EU}
\ee
where $U_\ell$ is given by \eqref{UL}, and $\ell$ is defined by \eqref{C1}. 
Thus, when the parameters $A,B,F,b$ of the trial function \eqref{A4} are set equal to the components of a fixed point of the system \eqref{Lagell}, the average field energy carried by the trial function coincides, up to the geometric factor $\pi^{3/2}/2$, with the value of the energy integral $H$ at that fixed point.


\section{Fixed points}  
\label{fixed} 

\subsection{  Dependence      on   frequency }

Having reduced the partial differential equation to a finite-dimensional system, we focus on its stationary points, which determine the parameters of the standing waves.

In the stationary case, where $\dot A=\dot B =\dot F=\dot b=0$, the differential Euler-Lagrange equations 
associated with the lagrangian \eqref{Lagell} become a system of four algebraic equations,
 \be
\left( \frac{\partial U_\ell}{\partial {\bf q}}  \right)_\ell=0.  \label{Z1}
 \ee
 Here the vector
 $\bf q$ comprises all four components of the fixed point: 
 \[
 {\bf q}= \left(\begin{array}{c} A \\ B \\ F \\ b \end{array} \right).
 \]

 The  brackets around the derivative in \eqref{Z1} indicate
 that  the parameter $\ell$ is held fixed during differentiation. 
It is convenient, however, to write this system in 
an equivalent form  
\be
\left(  \frac{\partial U_\omega}{ \partial {\bf q}}   \right)_\omega=0,
 \label{Uom}
\ee
where
 \[ 
 U_\omega=U-\frac{\omega^2}{2} I
 \] 
  and $I=I(A,F,b)$ is given by \eqref{phi}. The brackets in \eqref{Uom} imply that 
  the role of the  external parameter has returned to
 $\omega$.
  With the fixed-point components thus determined, $\ell$ can be recovered  as $\ell=\omega I$.

The system \eqref{Uom} reduces to two algebraic equations, \eqref{AA16} and \eqref{AA24} in Appendix \ref{AAA}. 
Those can be solved numerically or graphically (see Fig. \ref{Raph}).

The advantage of the system  \eqref{Uom} over \eqref{Z1} is that its
 roots    prove to be single-valued functions of $\omega$
--- while the system \eqref{Z1} has {\it two\/}  roots for each value of $\ell$.
Fig \ref{ABFb}  plots all four components of the fixed-point vector as well as the  invariant $\ell$. The energy integral \eqref{H}
 is shown  by the red curve in Fig \ref{En1}.

  \begin{figure}  
              \includegraphics[width=   \linewidth]{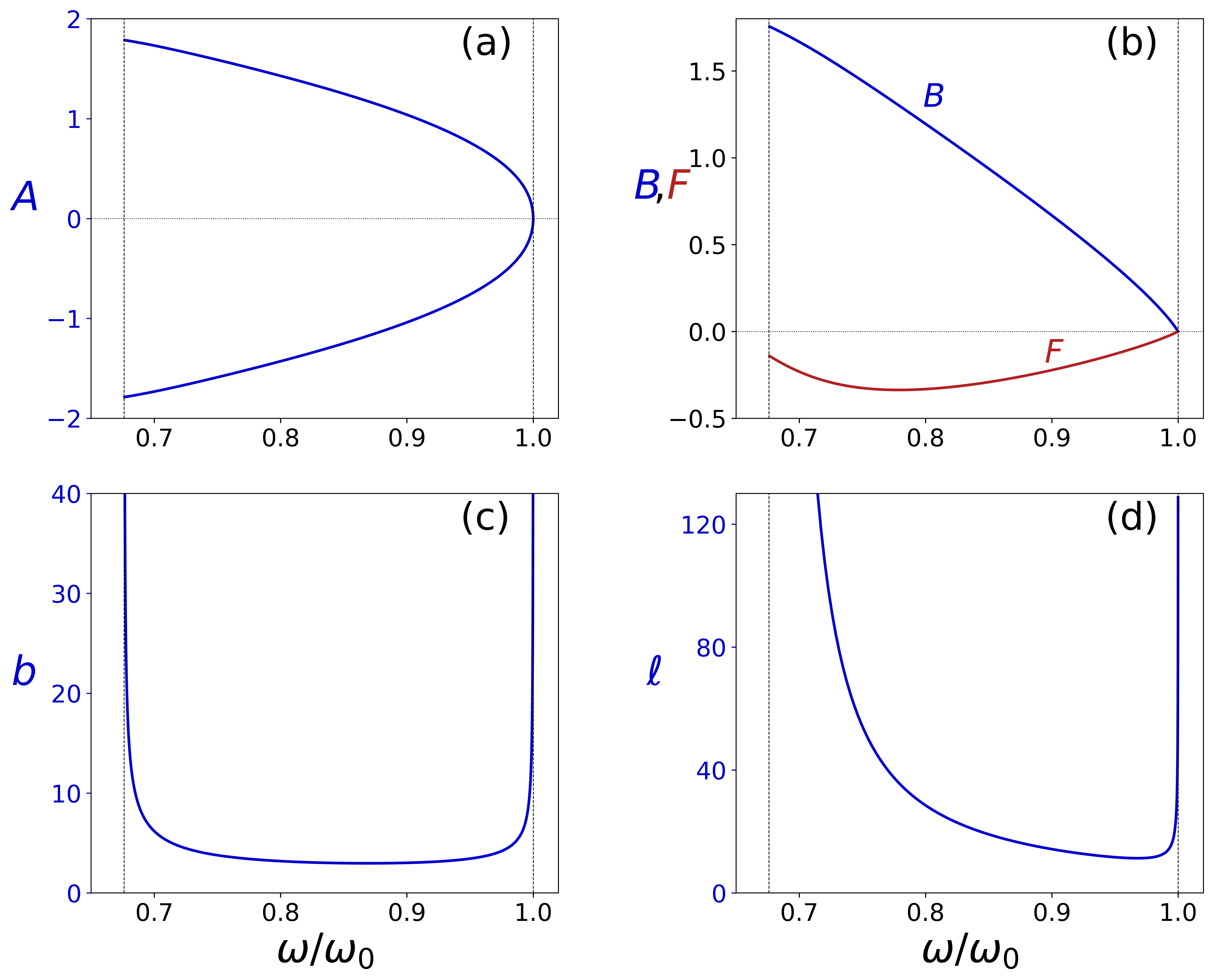}      
                     \caption{Components of the fixed point of the variational system \eqref{Lagell} and the adiabatic invariant as functions of $\omega$.
                      (a) amplitude of the first harmonic, $A(\omega)$; (b) amplitudes of the zeroth and second harmonics, $B(\omega)$ and $F(\omega)$; (c) spatial width of the trial function, $b(\omega)$; (d) adiabatic invariant, $\ell(\omega)$. As the dimensionless frequency $\omega/\omega_0$ approaches either $1$ from below or $0.676$ from above, the width $b(\omega)$ and hence the invariant $\ell(\omega)$ grow without bound.}
 \label{ABFb} 
\end{figure}

The range of existence of the fixed point is $\omega_* < \omega < \omega_0$, 
where
 $\omega_*/ \omega_0 = 0.67615959$.
As $\omega$ approaches $\omega_*$ from above, 
the width $b$ blows up as 
\be
b = \frac{    \, b_*    }   {\sqrt{     (\omega- \omega_*)   /\omega_0}} + ...,
\ee
where 
 $b_*= 1.1088567$, and 
  the ellipses stand for higher-order terms.

 Another noteworthy numerical observation concerns the minimum value of $\ell$ attained by the fixed-point solution. Specifically, $\ell$ reaches its minimum,
$\ell_{\mathrm{min}}=11.31$,
at $\omega_{\mathrm{min}}=0.96795 \,    \omega_0$. 
Remarkably, the same frequency also corresponds to a minimum of the energy \eqref{H}. This coincidence is explained 
in the next subsection.

 \subsection{Saddle-centre bifurcation} 
 
As we have pointed out in the previous subsection, 
 the fixed point ${\bf q}$ and, therefore,  invariant  $\ell$  are single-valued  functions of $\omega$.
  The derivative of the energy \eqref{H} with respect to this external parameter is given by
 \be
 \label{der}
    \frac{d H}{d \omega}=    \left(  \frac{\partial U_\ell}{\partial {\bf q}}  \right)_\ell \cdot \frac{ d {\bf q}}{d \omega}  + \frac{ \ell}{I }   \frac{d \ell}{d \omega}.       
 \ee
 In view of \eqref{Z1} and \eqref{C1}, we have
\be
 \label{Hl}
 \frac{d H}{d \omega}=  \omega   \frac{d \ell}{d \omega},
 \ee
 and so $dH/ d\omega=0$ whenever $d \ell/ d \omega=0$ (and the other way round). 
 This observation accounts for the coincidence of the minima of $H(\omega)$ and $\ell(\omega)$.

 \begin{figure}
       \includegraphics[width= 0.8 \linewidth]{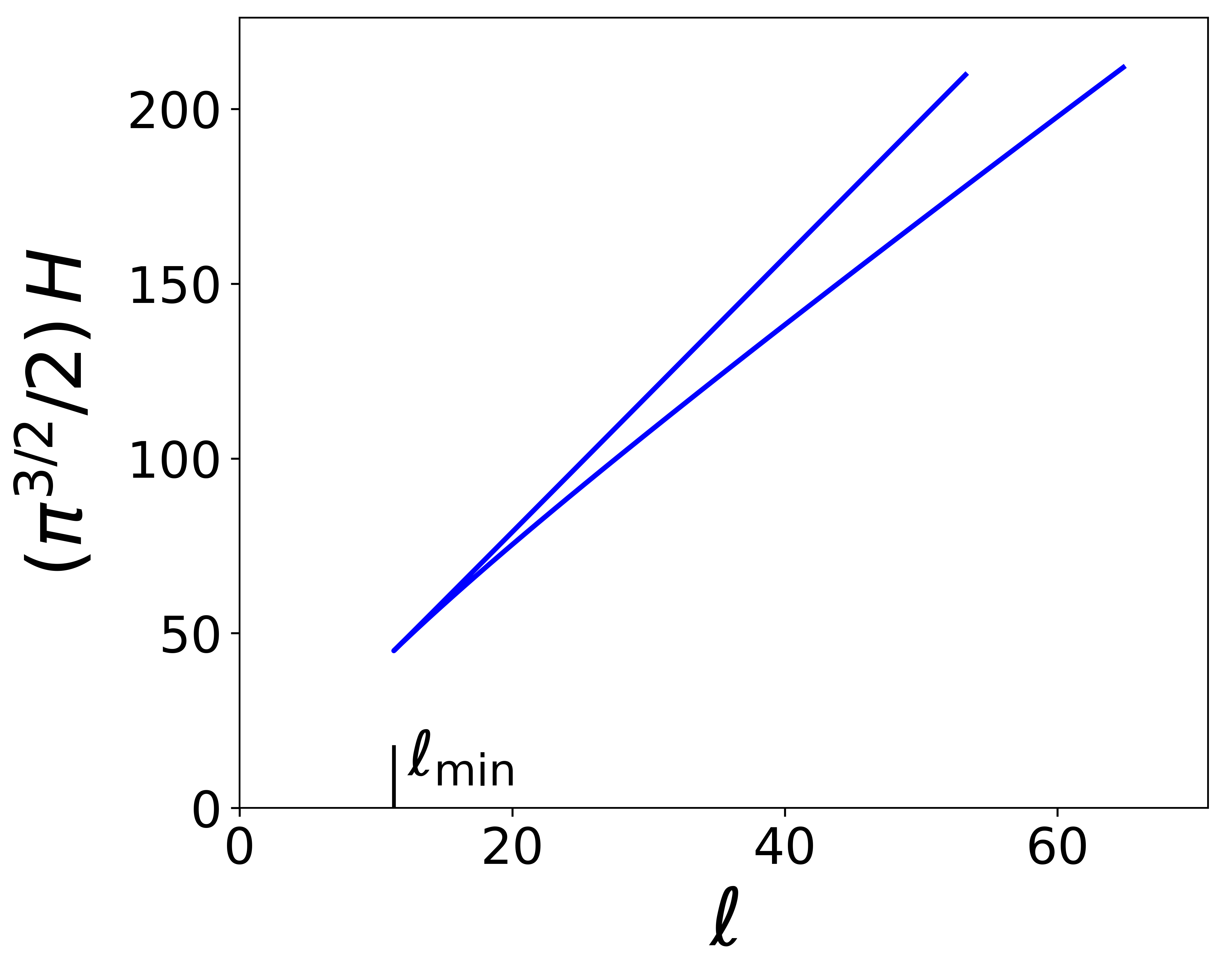}  
 \caption{Energy of the fixed point as a function of the invariant $\ell$. The bottom (stable)
 and the top (unstable)  branch correspond to $\omega< \omega_{\mathrm{min}}$ 
 and $\omega_{\mathrm{min}}< \omega < \omega_0$, respectively. 
 In this plot, the energy $U_\ell$ has been multiplied by $\pi^{3/2}/2$, as in \eqref{EU}, to be in line with Fig \ref{En1}.
 }
 \label{cusp} 
\end{figure}

It is worth noting that, unlike the invariant $\ell(\omega)$, the width function $b(\omega)$ attains its lowest value
 at $\omega=0.867761 \,\omega_0$,
 which differs from the frequency at which the energy is minimal.

When viewed as a function of $\ell$, the energy $H$ develops a turning point at $\ell=\ell_{\mathrm{min}}$ (Fig.~\ref{cusp}). Since
\[
\frac{dH}{d\ell}=   
\left( \frac{\partial U_\ell}{\partial {\bf q}}  \right)_\ell  \cdot \frac{d {\bf q}}{d \ell} + \frac{\ell}{I}
  =\omega,
\]
the slope remains  finite at the turning point, with
\[
\left.    \frac{dH}{d\ell}    \right|_{\ell=\ell_{\mathrm{min}}}=\omega_{\mathrm{min}}.
\]
Thus, the two branches of $H(\ell)$ meet with a finite slope, producing the cusp seen in Fig.~\ref{cusp}.

In a conservative system, such a turning point signals a saddle-centre bifurcation, at which the stability changes between the two branches. By Dirichlet's theorem, a stable equilibrium corresponds to a local minimum of the energy. Hence, the lower branch is stable, whereas the upper branch is unstable. We confirm this conclusion by a linearisation analysis,
after making a remark on the choice of collective coordinates.

\subsection{On the collective coordinate redundancy} 
\label{redundancy} 

One issue to settle before turning to the perturbation analysis is whether all five collective coordinates in the ansatz \eqref{A4} are necessary. The fundamental harmonic (with amplitude $A$) and the width ($b$) are natural characteristics of a localised standing wave, and the lazy phase ($\theta$) is essential as well \cite{BA}. We therefore examine the role of the remaining two coordinates, corresponding to the zeroth and second harmonics, with amplitudes $B$ and $F$, by considering the four-dimensional systems obtained by omitting each of them in turn. 
For details, see Appendix \ref{reduced}.

Setting $B=0$ in ansatz \eqref{A4} eliminates all fixed points of the resulting four-dimensional system. 
The zeroth harmonic is thus essential for the existence of the localised standing wave;
it originates from   the broken $\mathbb{Z}_2$ 
symmetry of the $\phi^4$ model.   In contrast, omitting the second harmonic leaves  the fixed point intact and produces only a marginal change in the energy--frequency relation. Indeed, the four- and five-variable approximations generate energy--frequency curves that are very close to each other, as illustrated by the black and red curves in Fig.~\ref{En1}.

\subsection{Stability}

To classify the stability of the fixed point, we linearise the Euler–Lagrange equations derived from the Lagrangian \eqref{Lagell}, keeping the parameter $\ell$ fixed.
Letting
\be
\delta {\bf q}= e^{( \lambda / \epsilon) \mathcal T_1} {\bf y},
\label{pert}
\ee
where ${\bf y}$ is a constant 4-vector,
gives a generalised eigenvalue problem
\be
M {\bf y} = - \lambda^2 P \,  {\bf y}.
\label{MP}
\ee
Here $M$ and $P$ are $4 \times 4$ symmetric real matrices, given 
in the Appendix \ref{BBB}, where the positive definiteness of $P$ is also established.
Note that the ansatz \eqref{pert} tacitly assumes the exponents $\lambda$ being of order $\epsilon$.

   Since $P$ is positive definite, 
all generalised eigenvalues $ \lambda^2$ are  real
and the exponents  form pairs of real or pure imaginary opposite values, $\lambda$ and $-\lambda$. 
The fixed point cannot undergo any 
 hamiltonian Hopf bifurcations and the standing wave cannot suffer oscillatory instabilities.

The generalised eigenvalues were computed numerically, with the results shown in Fig. \ref{EVs}. 
Of the four eigenvalues, three do not satisfy the assumption $\lambda^2=O(\epsilon^2)$ and must therefore be discarded. 
The remaining eigenvalue is also $O(1)$, except in the neighbourhood  of two values of $\omega$ where the corresponding branch $\lambda^2(\omega)$ crosses zero.

One such value is $\omega_0$. As $\omega\to\omega_0$, the generalised eigenvalue $\lambda^2$ approaches zero from above, implying the instability of the small-amplitude standing waves. This result agrees with the asymptotic analysis of Ref. \cite{ABBZ}, which establishes that, in the limit $\omega\to\omega_0$, the standing wave admits a growing perturbation of the form
\be
\delta \phi=  \left[ \mathcal F(\varepsilon \rho) + i \mathcal G (\varepsilon \rho )\right] e^{i\omega t+ \Lambda \varepsilon^2 t} + c.c..
\ee
Here $\Lambda=5.50$ is a real eigenvalue of the associated linearised operator, $\mathcal{F}(\rho)$ and $\mathcal{G}(\rho)$ are the components of the corresponding spherically symmetric eigenvector, and $\varepsilon^2=\omega_0-\omega$.
(We remind that $\epsilon$ and $\varepsilon$ are distinct small parameters; specifically, $\epsilon$ scales as $\varepsilon^2$.)

 The second zero crossing of the curve $\lambda^2(\omega)$ occurs at
$\omega_{\mathrm{min}}=0.96795\, \omega_0$. This is precisely the frequency at which both the energy \eqref{H} and the invariant $\ell$ attain their minima. 
As $\omega$ decreases through $\omega_{\mathrm{min}}$ and the fixed point passes from the upper to the lower branch of the $H(\ell)$ curve in Fig.~\ref{cusp},  the unstable eigenvalue crosses into the negative-$\lambda^2$ region.

In terms of the characteristic exponents $\lambda$, the transition corresponds to the collision of two opposite real 
 exponents and their subsequent departure along the positive and negative imaginary axes. 
 This is the saddle-centre bifurcation anticipated from the behaviour of  $H(\ell)$ that we described in the previous section.

Thus, the   eigenvalue analysis is  consistent with our two-time-scale assumption  only in the neighbourhoods of $\omega= \omega_0$ and $\omega_{\mathrm{min}}$. 
This, however,  suffices to classify the stability  of the standing wave in the entire region of validity of the variational approach.
(Namely, the wave is stable for $\omega \leq \omega_{\mathrm{min}}$
and
unstable in the interval $\omega_{\mathrm{min}} < \omega < \omega_0$.)
 Had there been any further stability-instability transitions,  they would have been detectable through additional zero crossings of the eigenvalue branch.

We should emphasise that the variational approach does not account for the radiation wing of the standing wave and therefore cannot detect instabilities associated with the wing. When the wing is thin, such instabilities are expected to have small growth rates and to become manifest only on long timescales. They are therefore of limited relevance to oscillons, which can be modelled by standing waves only over finite time intervals.

  \begin{figure} 
       \includegraphics[width= 0.49 \linewidth]{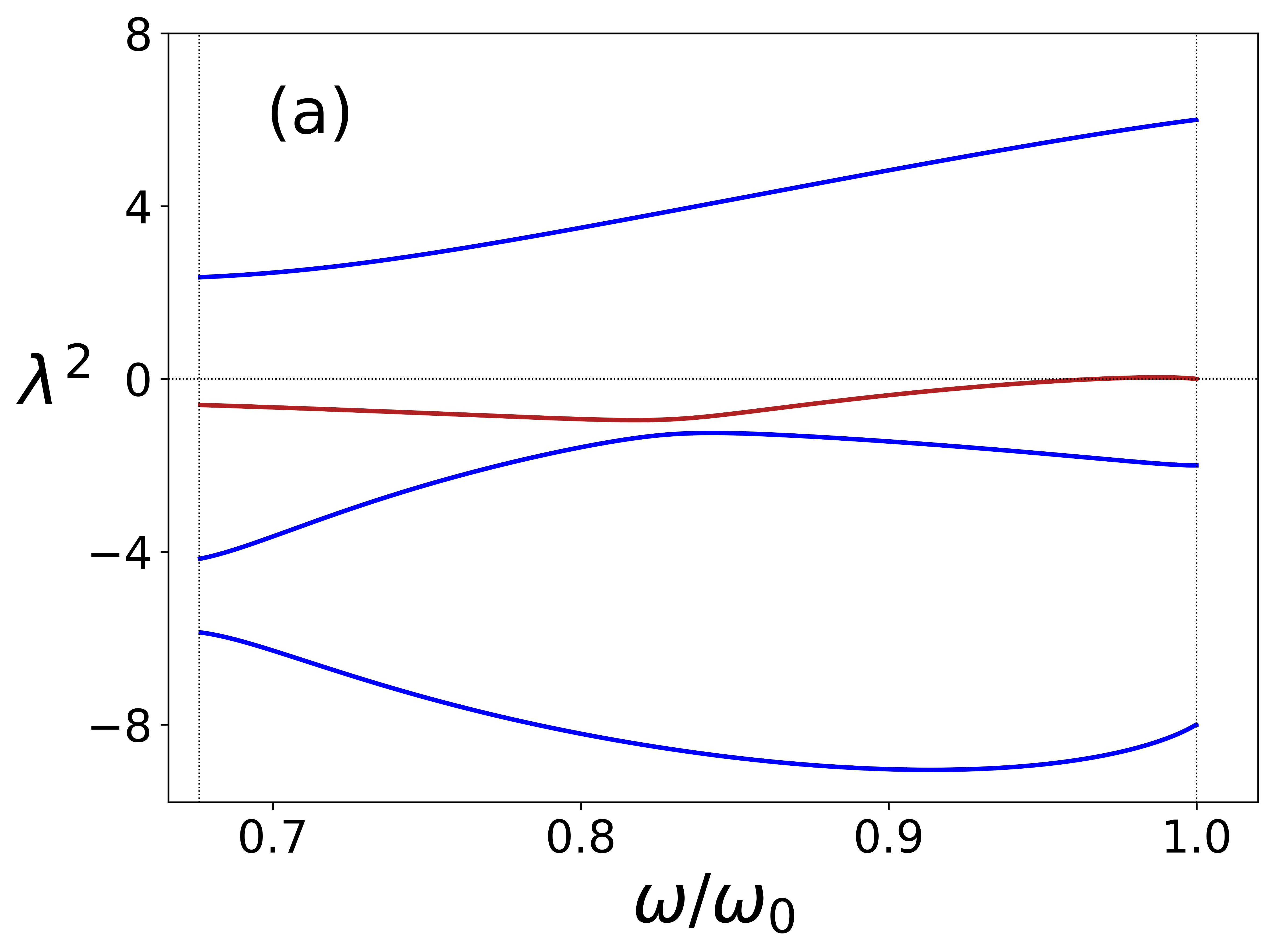}  
    \includegraphics[width=  0.49 \linewidth]{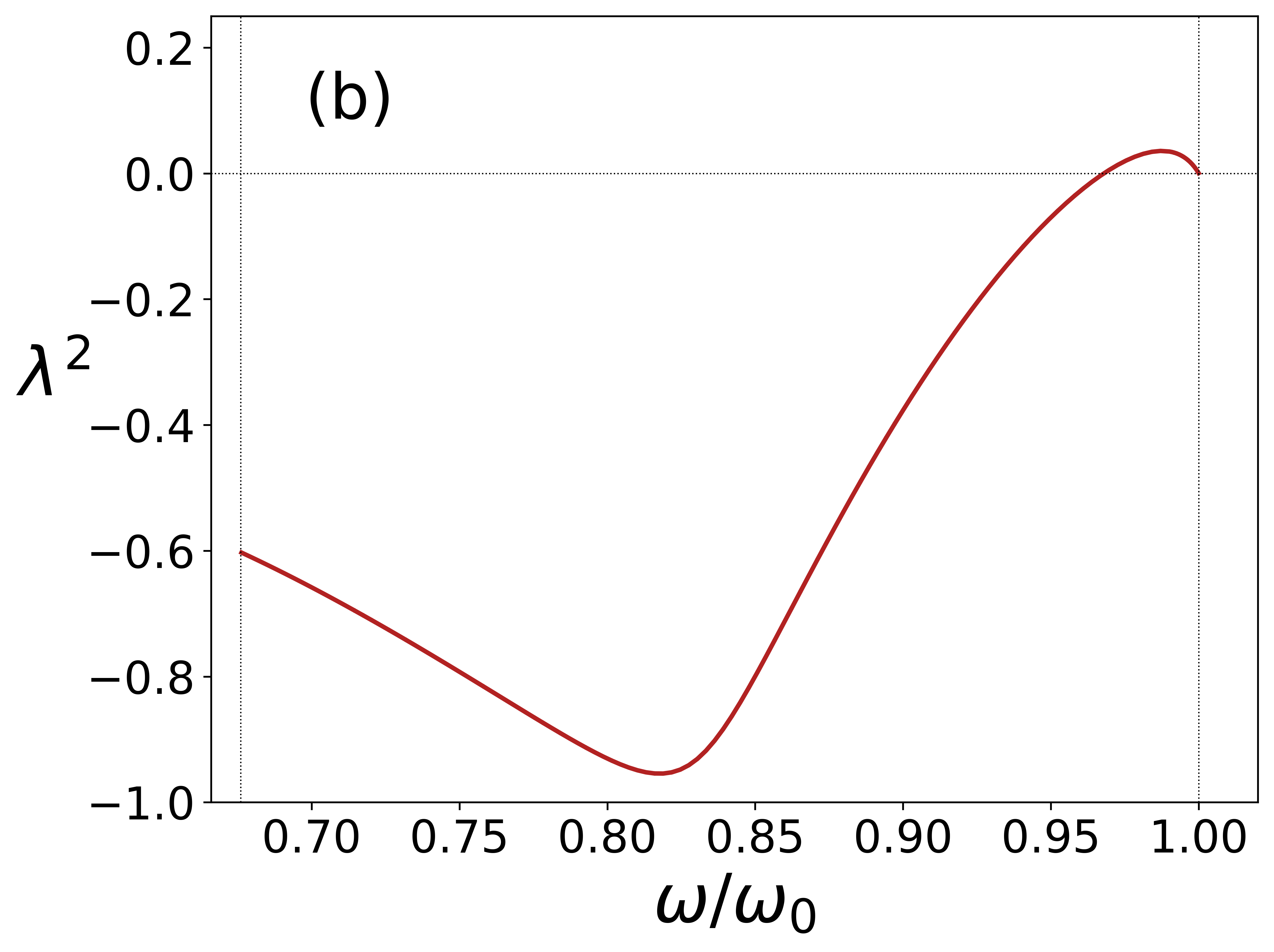}          
 \caption{Eigenvalues of the generalised eigenvalue problem \eqref{MP}. The eigenvalues shown by the blue curves in panel (a) have $\lambda=O(1)$ and are therefore inconsistent with the assumption $\lambda=O(\epsilon)$. Panel (b) shows a magnified view of the branch with the smallest $|\lambda|$, depicted in red in panel (a). This branch crosses $\lambda^2=0$ as $\omega$ is decreased through $\omega_{\mathrm{min}}=0.96795 \, \omega_0$, changing from positive to negative $\lambda^2$.}
 \label{EVs} 
\end{figure}

\section{Standing waves    --- numerical vs  variational}
\label{conclusions}

We have obtained the standing wave solutions of the $\phi^4$ equation as solutions of the periodic boundary-value problem \eqref{2D} with frequencies 
in the interval $     0.47711  <       \omega/ \omega_0 <1$. The results are represented in the form of an energy-frequency diagram in Fig \ref{En1}.

The dispersion curve in Fig.~\ref{En1} is similar to its one-dimensional analogue \cite{ABDD} in exhibiting a sequence of resonant spikes outlined by a smooth envelope.
In the one-dimensional case the smoothed energy decays monotonically and approaches zero as  $\omega \to \omega_0$.  
The behaviour of its three-dimensional counterpart at intermediate frequencies is similar — the smoothed energy decreases as the frequency grows.
However, unlike the 1D curve, the 
3D envelope features a minimum 
 (at a relatively high frequency, $\omega_{\mathrm{min}}=0.96732 \, \omega_0 $) and tends to infinity as $\omega \to \omega_0$.
 
 In both cases   the dispersion diagram is fragmented into two parts by a discontinuity at $\omega=\omega_0/2$. We attribute this fragmentation to the opening of a strong second-harmonic radiation channel for waves with $\omega>\omega_0/2$.

To get a theoretical handle on the numerical solutions, we have developed a variational approach to the standing waves. The method is a three-dimensional extension of the multiscale variational technique previously employed in the one-dimensional setting \cite{ABDD}. Our trial function is characterised by five collective coordinates: the amplitudes of the first,
  second and zeroth harmonics; an effective width of the wave core, and the ``lazy phase" (slowly varying  phase correction).
Note that  he radiation wing is not included in our variational ansatz.

   The five-parameter approximation accurately reproduces the envelope of the dispersion curve for $\omega/\omega_0 \gtrsim 0.85$. In particular, the variational energy--frequency curve has a single minimum at $\omega_{\mathrm{min}}/\omega_0=0.96795$, in close agreement with the minimum of the numerical envelope at $\omega/\omega_0=0.96732$.

The agreement deteriorates at lower frequencies: as  $\omega/\omega_0$  decreases from $0.85$, the variational curve deviates from the envelope of the numerically obtained energy--frequency relation.

A natural question for any finite-dimensional approximation is
whether the chosen set of collective coordinates  contains any redundant degrees of freedom.
We have verified that the  zero harmonic is an essential component of the variational ansatz.
The trial function missing  the static mode cannot capture any localised periodic oscillations.

By contrast,  omitting the localised second-harmonic mode  from the variational ansatz has no qualitative effect on the dispersion diagram.
 As seen in Fig.~\ref{En1}, the five- and four-mode approximations remain in close agreement. This suggests that the limitation of the variational approach stems not so much from its treatment of the wave core as from its inability to capture the coupling between the core and the radiation continuum.
This limitation becomes apparent at small $\omega$, where the amplitude of the radiation wing is no longer negligible. For $\omega < 0.85 \, \omega_0$, accurately reproducing the dispersion curve requires the  amplitude of radiation to be included in the 
 set of collective variables.

The variational analysis classifies the standing waves with $\omega>\omega_{\mathrm{min}}$ as unstable 
and predicts their stabilisation as $\omega$ decreases through $\omega_{\mathrm{min}}$. No further change of stability is detected  over the range where the two-time-scale approximation remains reliable.

The stabilisation of the variational fixed point  when $\omega$ drops below its energy-minimum value is consistent with the earlier numerical analysis \cite{ABBZ} of the boundary-value problem \eqref{2D}. (In particular,  the variational result for the value of $\omega_{\mathrm{min}}$
 is in excellent agreement with its  numerical counterpart.)  
Any further comparison with the stability properties of solutions of the boundary-value problem  is, however,  beyond the scope of the present study. 
We conjecture that  the monotonic instabilities  re-emerge below $\omega=0.85 \, \omega_0$, where the variational approximation no longer reproduces the numerical dispersion accurately.
 In addition, oscillatory instabilities are expected to arise from the radiation wing, whose amplitude becomes larger as $\omega$ decreases. 
Since our current ansatz contains no radiation component, such instabilities are necessarily absent from the variational description.

\section*{Acknowledgments}
 This research was supported by the bilateral collaborative 
 grant from  the Joint Institute  for Nuclear Research and National Research Foundation of South Africa (grant 120467).

\appendix
  
\section{Stationary points of the variational equations}
\label{AAA}

\subsection{Graphable  algebraic system} 

The aim of this Appendix is to reduce the system of four equations \eqref{Uom} to the intersection of two curves on the $(B,F)$ plane.  
  
  Before considering each  equation, it is convenient to
decompose the potential  with respect to  the powers of $b$:
\bea
U_\omega= \frac{3b}{4}  \left( \frac{A^2}{\sqrt 2} + B^2 + \frac{F^2}{2} \right) + b^3 W(A^2,B,F),      \nonumber   
\label{AA3} 
\eea
where the coefficient $W$ is further expanded in  powers of $A^2$:
\bea
W   =\frac14  \left( B^2 + \frac{F^2}{2} \right)
- \frac{B}{3 \sqrt 6} \left( B^2+ \frac32 F^2 \right)  \nonumber \\  -\frac{\omega^2}{4} F^2
  + \frac{1}{32 \sqrt 2}  \left( B^4+ 3B^2F^2 + \frac38 F^4 \right)
  \nonumber \\ 
  +  \frac{A^2}{2 \sqrt 2}   \left(1- \frac{\omega^2}{2} \right)
   - \mathcal G  \, A^2 
+  \frac{3}{128} A^4.
\label{AA12}
\eea
In \eqref{AA12} $\mathcal G$ stands for the following quadratic polynomial of $B$ and $F$:
\be 
\mathcal G(B,F)=    \frac38    \left( B + \frac{F}{2} \right)   - \frac{1}{4 \sqrt 6}    \left( B^2 +BF+ \frac{F^2}{2} \right). 
\nonumber 
\ee

Equation $\partial U_\omega / \partial A^2=0$ is linear in $A^2$; hence we can readily express $A^2$ in terms of the other variables:
\be
A^2 =  \frac{64}{3} \mathcal G  - 
8 \sqrt{2} \left(  b^{-2}   + \frac{2 -\omega^2}{3} \right).
\label{AA1} 
\ee
Turning to  $\partial U_\omega  / \partial B=0$ and $\partial U_\omega  / \partial F=0$, and 
making use of the decomposition \eqref{AA3}, we obtain
\be
 \frac32 b^{-2}  B + \frac{\partial W}{\partial B}=0               \label{AA4} 
\ee
and
\be
 \frac34 b^{-2}  F   +   \frac{  \partial W}{\partial F} =0,
\label{AA5}
\ee
respectively. 
Since $\partial W/\partial B$ and $\partial W/\partial F$ are linear in $A^2$ (the $A^4$ term drops out), substituting $A^2$ from \eqref{AA1} into \eqref{AA4} gives an equation linear in $b^{-2}$, whose solution is
\be
b^{-2}= \frac{f_\omega(B,F)}{  \mathcal D(B,F)}.
\label{AA6}
\ee
Here
\bea
f_\omega=
3B  - \left( 3 \sqrt 6    -  4B - 2F \right)      \nonumber \\
    \times  \left[
  \frac{8 \sqrt 6}{3} \mathcal G     \right. 
\left.
-  \frac{2}{\sqrt 3} (2- \omega^2) 
   \right]   
    \nonumber \\
        - \sqrt 6 \left( B^2+\frac{F^2}{2} \right)
+ \frac{3 \sqrt 2}{16} B (2B^2+ 3F^2)
\nonumber 
\eea
and
\be
\mathcal D= (8 \sqrt 3-9)B + 4 \sqrt 3 F - 18 \sqrt 2.
\nonumber 
\ee
Likewise, substituting \eqref{AA1} into \eqref{AA5} gives
\be
b^{-2}= \frac{ g_\omega(B,F)}  {  \tilde {\mathcal D} (B,F)},
\label{AA8}
\ee
with
\bea
g_\omega=
4   \sqrt 2   \left( F+B -\frac{3 \sqrt 6}{4}  \right) 
\left[ \frac83      \mathcal G  -
\frac{\sqrt 2}{3} (2- \omega^2)   \right]    \nonumber \\
+
 F \left[    \sqrt 3 (1-2\omega^2)    - \frac{4}{\sqrt 2} B + \frac{3\sqrt 3}{16 \sqrt 2} (F^2+4B^2) \right]
 \nonumber 
\eea
and 
\be
\tilde {\mathcal D} =  8 B + \left( 8-3 \sqrt 3 \right) F - 6 \sqrt 6.
\nonumber  
\ee
Finally, eliminating $b^{-2}$ between \eqref{AA6} and \eqref{AA8}, we obtain an equation for a curve on the $(B,F)$ plane:
\be 
f_\omega (B,F)  \tilde {\mathcal D} (B,F)  =g_\omega(B,F)  \mathcal D (B,F).
\label{AA16}
\ee

The remaining Euler--Lagrange equation, $\partial U_\omega/\partial b=0$, reads
\be
b^{-2} \left(\frac{A^2}{\sqrt 2} + B^2+\frac{F^2}{2} \right) + 4W  =0.
\label{AA10}
\ee
Using \eqref{AA12} for $W$ and \eqref{AA1} for $A^2$, we transform \eqref{AA10} into 
a second relation between $B$ and $F$:
\bea
\left[ (2-\omega^2) \mathcal D + 3 f_\omega \right]^2 + 3 \mathcal D p_\omega \left[ (2-\omega^2) \mathcal D + 3 f_\omega \right] 
\nonumber  \\   + 9 \mathcal D^2 q_\omega=0,
\label{AA24} 
\eea
with $p_\omega$ and $q_\omega$ given by
\be
p_\omega(B,F)=  \frac{B^2}{4}  +\frac{F^2}{8}      + \frac{8 \sqrt 2}{3} \mathcal G  + \frac{4}{3} (\omega^2-2), 
\nonumber 
\ee
and 
\bea
q_\omega (B,F)= 
\frac{\omega^2+1}{12}  \left(B^2+ \frac{F^2}{2} \right) -   \frac{\omega^2}{4}  F^2   \nonumber   \\
 - \frac{1}{3 \sqrt 6} B \left( B^2 +\frac32 F^2 \right)    
 + \frac{1}{32 \sqrt 2} \left(  B^4 + 3B^2F^2   \right.
 \nonumber    \\   \left. 
 + \frac38 F^4\right)     
    - \frac{ 16  \sqrt 2}{9 } \mathcal G \left(  3 \sqrt 2 \, \mathcal G + \omega^2 -2  \right).
 \nonumber 
\eea

The curves \eqref{AA16} and \eqref{AA24} are plotted in Fig.~\ref{Raph}, where the region of the $(B,F)$ plane in which $A^2>0$ and $b^{-2}>0$ simultaneously is  tinted. 
There is either one intersection of the two curves in that region or none, depending on whether $\omega$ lies above or below the threshold value $\omega_*= 0.67616 \, \omega_0$.
Hence the system \eqref{Uom} has a single fixed point for any $\omega$ between $\omega_*$ and $\omega_0$,
and no fixed points in the complementary  interval $(0, \omega_*)$.

  \begin{figure}
 \includegraphics[width= 0.49 \linewidth]{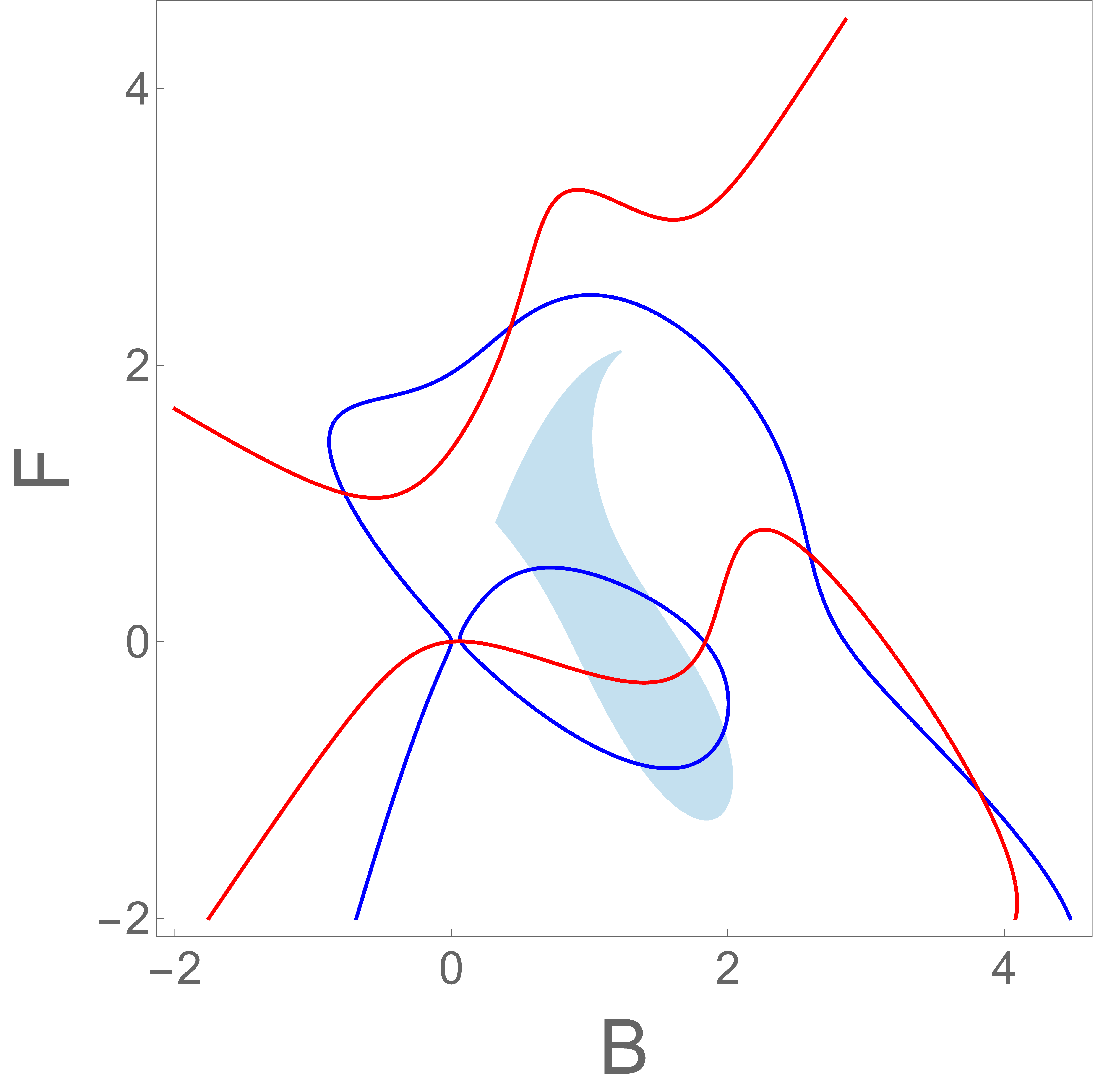}  
  \includegraphics[width= 0.49 \linewidth]{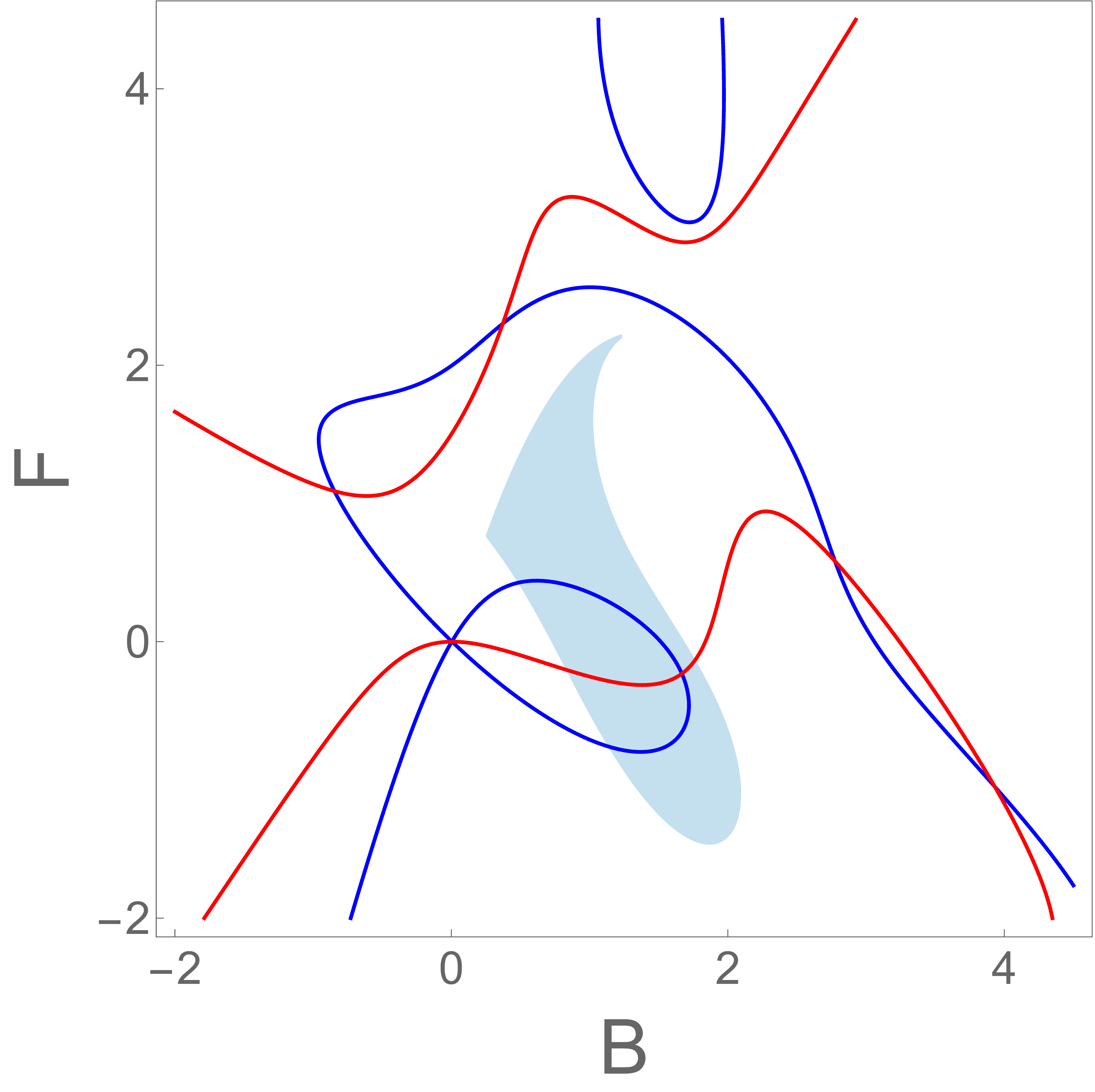}  
  \caption{Fixed points of the system \eqref{Uom} on the $(B,F)$ plane. The red and blue curves are given by \eqref{AA16} and \eqref{AA24}, respectively.
  The  admissible region where the inequalities $A^2>0$ and $b^{-2}>0$ 
  are satisfied simultaneously is tinted blue. 
  (a) $\omega/\omega_0= 0.65$ (no fixed points in the admissible region);
  (b) $\omega/\omega_0=0.70$ (a single fixed point). } 
 \label{Raph} 
\end{figure}

\subsection{Explicit fixed point in the limit $\omega \to \omega_0$}

As $\omega \to \omega_0$, the fixed point can be found in closed form.
Guided by \eqref{AA29}, we posit the scaling $A^2, B, F, b^{-2} \sim \varepsilon^2$, where $\varepsilon^2=\omega_0^2-\omega^2$.
At leading order, equation \eqref{AA6} then reduces to
\be 
\frac56 B + \frac{F}{2}= \sqrt 2 \left( b^{-2} + \frac{\varepsilon^2}{3} \right),
\nonumber   
\ee
while \eqref{AA8} yields
\be
B + F= \sqrt 2 \left( b^{-2} + \frac{\varepsilon^2}{3} \right).
\nonumber \\   \label{AA26}
\ee
Solving this pair for $B$ and $F$ in terms of $b^{-2}$, we find
\be
F= -\frac{1}{\sqrt 2} \left( b^{-2} + \frac{\varepsilon^2}{3} \right),
\quad 
B=-3F.
\label{AA27} 
\ee

To fix the coefficient relating $b^{-2}$ to $\varepsilon^2$, we turn to \eqref{AA10}, which at order $\varepsilon^4$ reads
\bea
-5B^2 - 3F^2 - 6BF - \frac43 \varepsilon^4 + 4 b^{-4} - \frac83 \varepsilon^2 b^{-2} \nonumber \\
+ \frac{8}{\sqrt 2} \left( B + \frac{F}{2} \right) (b^{-2} + \varepsilon^2)=0.
\nonumber  
\eea
Inserting \eqref{AA27} turns this into a quadratic equation for $b^{-2}$, whose negative root, $-\varepsilon^2/3$, is inadmissible; the positive root gives
\be
b^{-2} = \varepsilon^2.
\nonumber   
\ee
The amplitudes of the zeroth and second harmonics then follow at once from \eqref{AA27}:
\be 
B= 2 \sqrt 2 \varepsilon^2, \quad F=- \frac{2 \sqrt 2}{3} \varepsilon^2,
\nonumber    
\ee
and \eqref{AA1} supplies the squared amplitude of the main harmonic:
\be
A^2= \frac{8 \sqrt 2}{3} \varepsilon^2.
\nonumber   
\ee

We note that  $B=\tfrac34 A^2$ and $F=-\tfrac14 A^2$, in agreement with the relations between the amplitudes of the first, zeroth and second harmonics in the expansion \eqref{AA29}.

\subsection{Reduced collective coordinate sets} 
\label{reduced}

Here we examine   ``curtailed" variational approximations corresponding to
setting $B=0$  or $F=0$ in  the trial function  \eqref{A4}.

Let, first, $B=0$ (while $A \neq 0$). The equation $\partial U_\omega / \partial A^2=0$ gives
\be
\label{Q1}
 F= \frac{A^2}{4} + \frac{\sqrt 6}{9}   F^2   + 2 \sqrt 2  \left(   b^{-2}  + \frac{2-\omega^2}{3}  \right),
\ee
whence $F>0$. 
Eliminating  frequency offset $2-\omega^2$ between \eqref{Q1} 
and equation $\partial U_\omega/ \partial F=0$ gives
\be
b^{-2}= -    \frac{1}{72} v(A,F),
\label{Q2}
\ee
where
\be
v= A^2 \left[ \frac{18}{F}  +  \sqrt 6 (3 \sqrt 3-4) \right]      +u(F),
\label{Q3}
\ee
and $u(F)$ is a quadratic polynomial with a positive leading coefficient and negative discriminant:
\[
u= \left( 8 \sqrt 3 - \frac{9}{2 \sqrt 2}  \right) F^2     -  36 \sqrt 2 F + 72.
\]

Since $F>0$ while  $u(F) >0$ for any $F$,
 both terms in the right-hand side of \eqref{Q3} are positive.
Equation \eqref{Q2} implies then $b^{-2}<0$. This is inconsistent with the ansatz \eqref{A4} which introduces $b$ as a \emph{real}  width of the trial waveform. 
We conclude that 
a nonvanishing zeroth harmonic is  an essential component of the standing wave.

To examine the role of 
the second harmonic,  we set  $F=0$  in    the potential \eqref{AA3}.
Equations 
$\partial U_\omega  / \partial A^2=0$ and $\partial U_\omega / \partial B=0$ constitute a linear system for  $A^2$ and $b^{-2}$
(where we assume that neither of these two unknowns is zero). 
The solution is 
\begin{align}
b^{-2}=   r_\omega(B),         \label{X1}  \\
A^2  =
8 \sqrt 2 \left[\frac{B}{\sqrt 2} 
-\frac{B^2}{\sqrt{27}}
-\frac{2-\omega^2}{3}
-   r_\omega(B) \right],   
  \label{X2}
\end{align}   
where  $r_\omega$ stands for the  ratio
\[
r_\omega(B)
\equiv
\frac{\mathcal N_\omega(B)}{3\,\mathcal D_0(B)} 
\]
with
\begin{align*}
\mathcal N_\omega(B)
=
15 \sqrt6\,B^2
+\left( \frac{9}{4\sqrt2} -8\right)B^3 \\
+  18    \sqrt2    \left( 2- \omega^2\right)
-   B\left[   45+8 \sqrt 3(2-\omega^2) \right]
\end{align*}
 and 
  \[
\mathcal D_0(B)   =\left(8\sqrt3 -9    \right)B         - 18\sqrt2.
\] 

The remaining Euler-Lagrange equation,
$\partial U_\omega /\partial b=0$, takes the form   
\[
b^{-2}     \left( \frac{A^2}{\sqrt2}     +    B^2   \right) 
+    \left.   4W  \right|_{F=0}
=0 .
\]
Upon substituting  $b^{-2}$  and   $A^2$  from \eqref{X1} and \eqref{X2}, this equation  reduces to
\[
\frac{B\,\mathcal P_\omega(B)}
{48\,\mathcal D_0^{\,2}(B)}
=0,
\]
where $\mathcal P_\omega$ is a quintic polynomial  in $B$:
\begin{align}
\mathcal P_\omega(B)
=
\left(522+111\sqrt2-1024\sqrt3+144\sqrt6\right)B^5
\nonumber\\
   -720\left(-16\sqrt2+2\sqrt3+3\sqrt6\right)B^4
\nonumber\\
    -24B^3\Big[
20-81\sqrt2+1128\sqrt3-64\sqrt6
\nonumber\\
    \hspace{22mm}
+2\left(-192-9\sqrt2+48\sqrt3+16\sqrt6\right)\omega^2
\Big]
\nonumber\\
-96B^2\Big[
72-7\sqrt2-444\sqrt6
+\left(252\sqrt6      -36-199\sqrt2                  \right)\omega^2
\Big]
\nonumber\\
   -192B\Big[
639-32\sqrt3
-48\left(6+\sqrt3\right)\omega^2
+\left(32\sqrt3       -27  \right)\omega^4
\Big]
\nonumber\\
+13824\sqrt2\left(2-3\omega^2+\omega^4\right).
\label{qui}
\end{align}

 Depending on $\omega$, the polynomial $\mathcal P_\omega(B)$ has either one or three real roots. 
 We evaluate these roots numerically for $0<\omega<\omega_0$ and, for each root, obtain the corresponding values of $b^{-2}$ and $A^2$  from \eqref{X1} and \eqref{X2}. 
Of the resulting branches, only one satisfies both constraints, $r_\omega(B)>0$ and $A^2(B)>0$, simultaneously. It exists for $0.6737\,\omega_0<\omega<\omega_0$;
  see Fig.~\ref{quintic}.
  As $\omega$ decreases to the lower limit of this interval, 
  the ratio $r_\omega$ (that is, $b^{-2}$)  approaches zero, so the width of the trial waveform  grows without bound and the branch terminates.
  Note that the existence  threshold  is very close to the value $\omega_*=0.67616\,\omega_0$ found for the  ansatz including all five collective variables.

  Substituting all components of the fixed point into the expression \eqref{EU}, we plot the energy carried by the trial function with $F=0$ 
  (black curve in Fig.~\ref{En1}). This curve differs only slightly from the one obtained with the full trial function (red curve), so the second harmonic, unlike the zeroth, is not indispensable for the existence of the standing wave: it only refines the energy of the solution.

  \begin{figure}
 \includegraphics[width= 0.9 \linewidth]{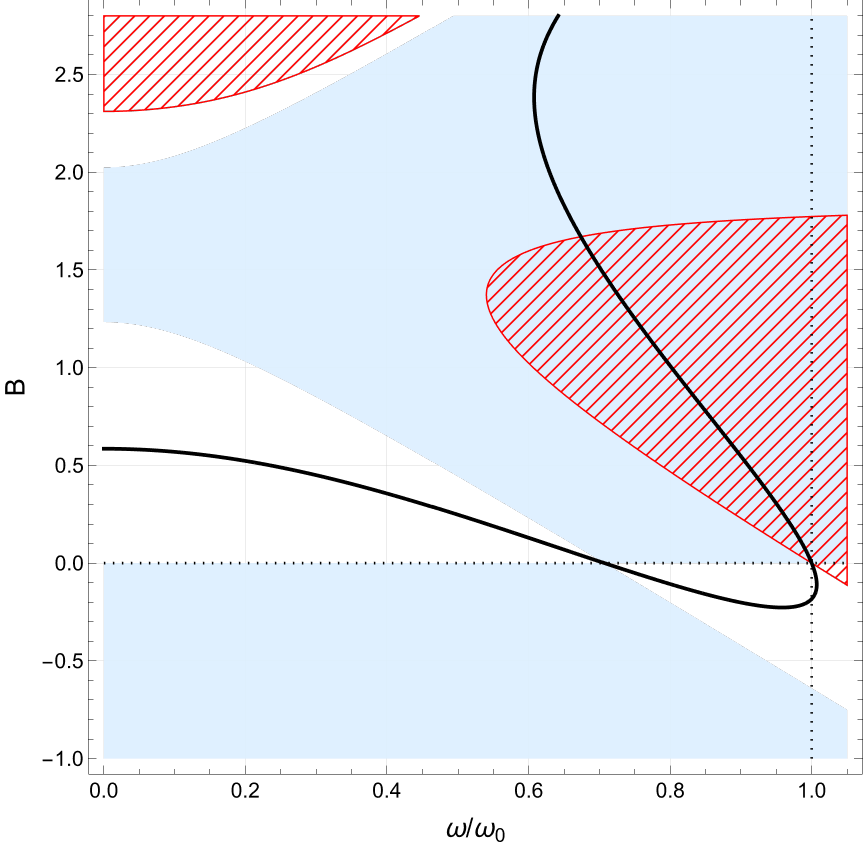}  
   \caption{ Black curve: real roots of the quintic polynomial \eqref{qui} as functions of  the frequency. The red-hatched and  blue-tinted
  regions correspond to $b^{-2} >0$ and $A^2>0$, respectively; only one root of $\mathcal P_\omega(B)$ lies within their overlap.  
  The admissible segment of the black curve extends over \(0.6737 \leq \omega/\omega_0 \leq 1\), with the left endpoint corresponding to \(b^{-2}=0\).  } 
 \label{quintic} 
\end{figure}   
 
  \section{Matrix of linearised variational equations} 
 \label{BBB} 

 The matrix elements $M_{ij}=M_{ji}$ in equation \eqref{MP} are given by
\begin{align}
M_{11}={}&\frac{3b}{2}+b^3
-\frac{3Bb^3}{2\sqrt2}
-\frac{3Fb^3}{4\sqrt2}
+\frac{B^2b^3}{2\sqrt3}
+\frac{BFb^3}{2\sqrt3}
\nonumber\\
&+\frac{F^2b^3}{4\sqrt3}
+\frac{9A^2b^3}{16\sqrt2}
+\frac{\omega^2b^3\left(3\sqrt2A^2-2F^2\right)}
{2\left(\sqrt2A^2+2F^2\right)}
\nonumber\\[0.5em]
M_{12}={}&
-\frac{3Ab^3}{2\sqrt2}
+\frac{ABb^3}{\sqrt3}
+\frac{AFb^3}{2\sqrt3}
\nonumber\\[0.5em]
M_{13}={}&
-\frac{3 \sqrt{2} Ab^3}{8}
+\frac{ABb^3}{2\sqrt3}
+\frac{AFb^3}{2\sqrt3}
+\frac{4AF\omega^2b^3}
{\sqrt2A^2+2F^2}
\nonumber\\[0.5em]
M_{14}={}&
\frac{3A}{2}
+3Ab^2
-\frac{9ABb^2}{2\sqrt2}
-\frac{9AFb^2}{4\sqrt2}
+\frac{\sqrt3AB^2b^2}{2}
\nonumber\\
&+\frac{\sqrt3ABFb^2}{2}
+\frac{\sqrt3AF^2b^2}{4}
+\frac{9A^3b^2}{16\sqrt2}
+\frac{3A\omega^2b^2}{2}
\nonumber
\end{align}
\begin{align}
M_{22}={}&
\frac{3b}{\sqrt2}
+\frac{b^3}{\sqrt2}
-\frac{2Bb^3}{\sqrt3}
+\frac{3B^2b^3}{8}
+\frac{3F^2b^3}{16}
+\frac{A^2b^3}{2\sqrt3}
\nonumber\\[0.5em]
M_{23}={}&
-\frac{Fb^3}{\sqrt3}
+\frac{3BFb^3}{8}
+\frac{A^2b^3}{4\sqrt3}
\nonumber\\[0.5em]
M_{24}={}&
\frac{3B}{\sqrt2}
+\frac{3Bb^2}{\sqrt2}
-\sqrt3B^2b^2
-\frac{\sqrt3F^2b^2}{2}
+\frac{3B^3b^2}{8}
\nonumber\\
&+\frac{9BF^2b^2}{16}
-\frac{9A^2b^2}{4\sqrt2}
+\frac{\sqrt3A^2Bb^2}{2}
+\frac{\sqrt3A^2Fb^2}{4}
\nonumber
\end{align}
\begin{align}
M_{33}={}&
\frac{3b}{2\sqrt2}
+\frac{b^3}{2\sqrt2}
-\frac{Bb^3}{\sqrt3}
+\frac{3B^2b^3}{16}
+\frac{9F^2b^3}{64}
+\frac{A^2b^3}{4\sqrt3}
\nonumber\\
&-\frac{\omega^2b^3\left(A^2-3\sqrt2F^2\right)}
{\sqrt2A^2+2F^2}
\nonumber\\[0.5em]
M_{34}={}&
\frac{3F}{2\sqrt2}
+\frac{3Fb^2}{2\sqrt2}
-\sqrt3BFb^2
+\frac{9F^3b^2}{64}
+\frac{9B^2Fb^2}{16}
\nonumber\\
&-\frac{9A^2b^2}{8\sqrt2}
+\frac{\sqrt3A^2Bb^2}{4}
+\frac{\sqrt3A^2Fb^2}{4}
+\frac{3\sqrt2F\omega^2b^2}{2}
\nonumber\\[0.5em]
M_{44}={}&
3A^2b
-\frac{9A^2Bb}{2\sqrt2}
-\frac{9A^2Fb}{4\sqrt2}
+\frac{\sqrt3A^2B^2b}{2}
+\frac{\sqrt3A^2BFb}{2}
\nonumber\\
&+\frac{\sqrt3A^2F^2b}{4}
+\frac{9A^4b}{32\sqrt2}
+\frac{3B^2b}{\sqrt2}
-\frac{2B^3b}{\sqrt3}
+\frac{3B^4b}{16}
\nonumber\\
&-\sqrt3BF^2b
+\frac{9B^2F^2b}{16}
+\frac{3F^2b}{2\sqrt2}
+\frac{9F^4b}{128}
+3A^2\omega^2 b \nonumber\\&+ 3\sqrt{2}\,F^2\omega^2 b
\nonumber
\end{align}
Here  $A, B, F$ and $b$ are components of the root of the system \eqref{Uom}.

The $P$ matrix in equation \eqref{MP}  has the form 
\be
P = \left( 
\begin{array}{cccc} 
\frac{b^3}{2} & 0 & 0 & \frac{3 A b^2}{4} \\
0 & \frac{b^3}{2 \sqrt{2}} & 0 & \frac{3 b^2 B}{4 \sqrt{2}} \\
0 & 0 & \frac{b^3}{4 \sqrt{2}}  & \frac{3 b^2 F}{8 \sqrt{2}} \\
\frac{3 A b^2}{4} & \frac{3 b^2 B}{4 \sqrt{2}}& \frac{3 b^2 F}{8 \sqrt{2}} & P_{44}
\end{array}
\right),
\nonumber 
\ee
where
\[
P_{44}= \frac{15b}{8} \Big(A^2+\frac{B^2}{\sqrt 2}+ \frac{F^2}{2 \sqrt 2}\Big).
\]

The determinant of $P$ is given by
\be
\nonumber   
\det P= \frac{3}{512} b^{10} \left[ 4 A^2+\sqrt{2} \left(2 B^2+F^2\right)\right].
\ee
Since the determinants of all leading principal minors are positive, 
  Sylvester's criterion guarantees that  $P$ is positive definite.

\end{document}